\documentclass[aps,prb,preprint,superscriptaddress,showpacs,endfloats*]{revtex4}

\usepackage{amsmath,amsfonts,bm,graphicx}

\newcommand{\eps}{\varepsilon}
\newcommand{\om}{\omega}

\newcommand{\llangle}{\langle\!\langle}
\newcommand{\rrangle}{\rangle\!\rangle}
\newcommand{\rlangle}{\rrangle\!\llangle}
\newcommand{\LL}{{\mathcal L}}
\newcommand{\PP}{{\mathcal P}}
\newcommand{\QQ}{{\mathcal Q}}

\begin{document}

\title{Current and current fluctuations in quantum shuttles}

\author{Antti--Pekka Jauho}
\email{antti@mic.dtu.dk}

\affiliation{MIC - Department of Micro and Nanotechnology,
             Technical University of Denmark,
             DTU - Building 345east,
             DK-2800 Kongens Lyngby, Denmark}

\author{Christian Flindt}
\email{cf@mic.dtu.dk}

\affiliation{MIC - Department of Micro and Nanotechnology,
             Technical University of Denmark,
             DTU - Building 345east,
             DK-2800 Kongens Lyngby, Denmark}

\author{Tom\'a\v s Novotn\'y}
\email{tno@mic.dtu.dk}

\affiliation{MIC - Department of Micro and Nanotechnology,
             Technical University of Denmark,
             DTU - Building 345east,
             DK-2800 Kongens Lyngby, Denmark}

\affiliation{Department of Electronic Structures,
     Faculty of Mathematics and Physics, Charles University,
     Ke Karlovu 5, 121 16 Prague, Czech Republic}

\author{Andrea Donarini}
\email{ad@mic.dtu.dk}

\affiliation{MIC - Department of Micro and Nanotechnology,
             Technical University of Denmark,
             DTU - Building 345east,
             DK-2800 Kongens Lyngby, Denmark}

\date{\today}

\begin{abstract}
We review the properties of electron shuttles, i.e.
nanoelectromechanical devices that transport electrons one-by-one
by utilizing a combination of electronic and mechanical degrees of
freedom.  We focus on the extreme quantum limit, where the
mechanical motion is quantized.  We introduce the main theoretical
tools needed for the analysis, e.g. generalized master equations
and Wigner functions, and we outline the methods how the resulting
large numerical problems can be handled.  Illustrative results are
given for current, noise, and full counting statistics for a
number of model systems. Throughout the review we focus on the
physics behind the various approximations, and some simple
examples are given to illustrate the theoretical concepts. We also
comment on the experimental situation.

\end{abstract}


\maketitle

\section{Introduction}

As advances in technology push the size of the electronic
components towards the nanometer scale, the well-established
technology of Micorelectromechanical Systems (MEMS) begins to
acquire quantum features.  This trend signals the birth of a new
research field, Nanoelectromechanical Systems (NEMS). Despite of
the infancy of NEMS, a large literature is already available on
it, and for a broad overview the interested reader is referred to
recent review articles \cite{cra-sci-00,cleland,ble-phr-04}. The
review at hand has a more restricted scope: it is devoted to a
theoretical analysis of a very specific NEMS device, the electron
shuttle.  By presenting a detailed study of such an idealized
model system we hope to be able to illustrate the basic physics
and conceptual problems that need to be understood before a
general theory of NEMS can be developed.

The electron shuttle, originally introduced by Gorelik {\it et
al.} in 1998\cite{gor-prl-98}, consists of a movable nanoscopic
grain which is coupled via tunnel barriers to source and drain
electrodes. A subtle combination of quantum transport and the
mechanical degrees of freedom  play an essential role for the
functionality of the device. The device can exhibit a ``phase
transition": when a control parameter is tuned (this could be the
damping of the oscillator), at a certain threshold value the
system enters a new transport regime, the shuttle regime, where
charge is transported in an orderly fashion (i.e. essentially a
fixed number charges per mechanical oscillation cycle). (This
phenomenon is particularly pronounced if the device operates in
the strong Coulomb blockade regime, when only one excess charge at
a time is allowed in the movable part.) In the original suggestion
of Ref. \onlinecite{gor-prl-98} the motion of the movable grain
was treated macroscopically, i.e., with Newton's equation of
motion. In our group we have been asking questions of what happens
when the movable body is so light that also its motion becomes
quantized. For example, does the shuttle transition persist in the
quantum regime? Another interesting question concerns the
interpretation of measurements of NEMS devices, possibly
exhibiting shuttling. As it turns out, a measurement of the
stationary IV-characteristics does not always yield enough
information to uniquely identify the underlying microscopic charge
transport mechanism. A point in case is  the C$_{60}$ SET
experiment by Park et al.\cite{par-nat-00} where two alternative
interpretations, namely incoherent phonon assisted
tunneling\cite{boe-epl-01, mcc-prb-03, fle-prb-03, braig-prb-03}
or shuttling,\cite{gor-prl-98, fed-epl-02} are plausible. The {\it
current noise} provides another important characteristics,
supplementary to the mean
current.\cite{kogan,bla-phr-00,bee-pht-03} The Fano factor which
characterizes the degree of correlation between charge transport
events is a powerful diagnostic tool  to distinguish
between various transport mechanisms. 
Therefore, studies of the current noise in NEMS have become an
active field of
research.\cite{mit-preprint-03,pis-prb-04,arm-preprint-04,
cht-preprint-04,isa-epl-04,nov-prl-04,bla-prl-04} Based on studies
where the mechanical system is treated classically, one expects a
giant enhancement of noise at the shuttling transition
\cite{isa-epl-04}, and we want to investigate whether this also
occurs in the quantum case. One can go even further along this
line: why stop at the current noise, which is the second cumulant
of the {\it full counting statics} (FCS)? Surely FCS (which
equals the probability $P_n(t)$ of $n$ electrons being collected,
say, in the right lead at time $t$) will contain even more
detailed information. Consequently, FCS in mesoscopic devices has
been a very active research field for some time now, and its
importance and relevance have been underlined by a recent {\it
measurement} of the third cumulant \cite{reu-prl-03}. Thus, FCS
for NEMS is obviously an interesting issue.  We are aware of a
recent calculation for a classical, driven shuttle
\cite{pis-prb-04}, and at the end of this paper we illustrate some
of our own very recent generalizations of these concepts to the
quantum regime using a simple example; a full description  will be
given elsewhere \cite{fli-epl-04}.

We have investigated quantum shuttles in several recent papers,
\cite{nov-prl-03,modena,nov-prl-04,fli-prb-04,fli-epl-04,andrea}
where the full technical details can be found; the purpose of the
present paper is to introduce some of the basic issues to a more
general reader. Several other groups have also recently studied
quantum shuttles, see e.g. Refs.
\onlinecite{she-jpc-03,fed-prb-03,fed-prl-04,smi-prb-04}. The
paper is organized as follows. In Section 2 we introduce the
models of  quantum shuttles employed in our work. 
The total Hamiltonian consisting of the ``system" (both mechanical
and electronic degrees of freedom of the quantum dot array), the
leads, and a generic heat bath is used to illustrate the
derivation of a  Markovian generalized master equation (GME) which
is the starting point of the theoretical analysis. Along the way
from the Hamiltonian to the generalized master equation we
identify the necessary assumptions, and point out several issues
of potential importance not addressed so far within the field of
NEMS.

In Section 3 we discuss the calculation of the current and the
zero-frequency component of the current noise spectrum for a  NEMS
device described by a GME. We also give a qualitative discussion
of the required numerical calculations, which are highly
non-trivial due to the large dimensions of the involved matrices.
Finally, some examples of our numerical results are displayed.
Section 4 is devoted to an elementary discussion of the
calculation of the full counting statistics, and a brief review of
the experimental status.

\section{The Model}

\begin{figure}
  \centering
 \includegraphics[width=\textwidth]{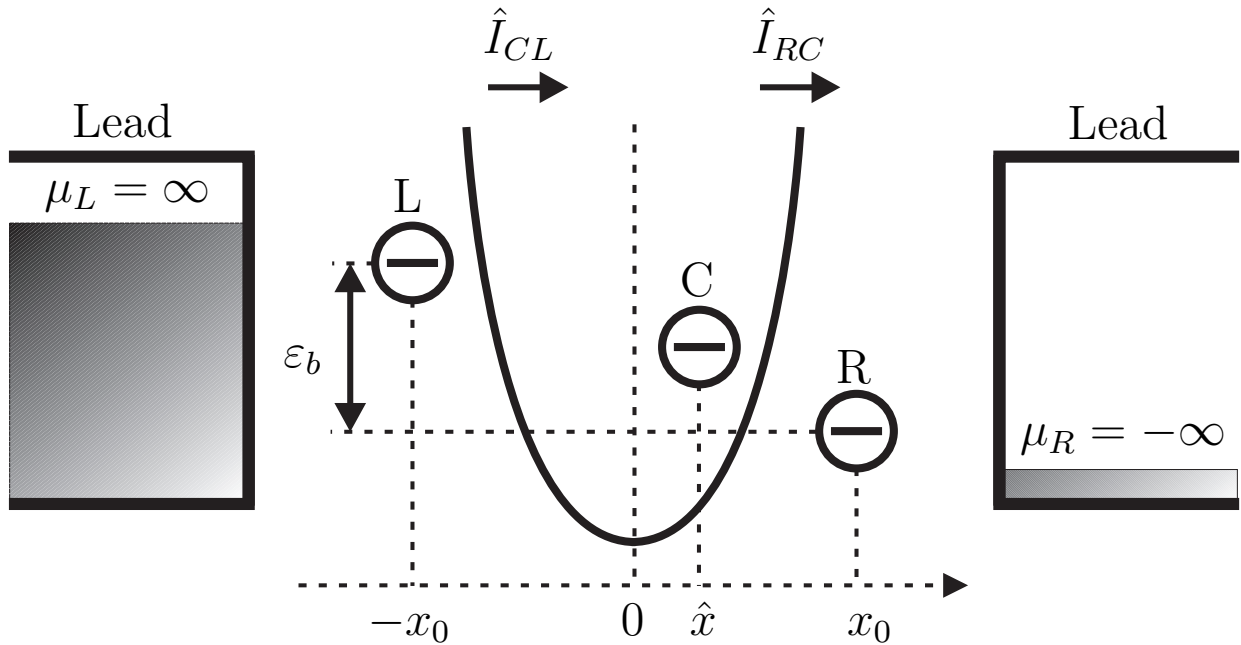}
  \caption{Schematic picture of a three-dot system, introduced by Armour and MacKinnon
  \cite{arm-prb-02}. The outer
  dots are fixed --- the left one (L) at the position $-x_0$ and the right one
  (R) at $x_0$, while the central one (C) can move (position $\hat{x}$) in
  a harmonic confining potential. It also interacts with a heat bath causing damping
  and thermal noise. The outer dots whose respective energy levels are
  de-aligned by the device bias ($\eps_b$) are coupled to the full or empty electronic
  reservoirs (leads), respectively. The current flows within the system due to tunneling
  between the left and central dot and the central and right dot.
  (Reproduced from Ref. \onlinecite{nov-prl-03}).
}\label{setup}
\end{figure}

At least two model systems have been considered in the literature.
Gorelik {\it et al.} \cite{gor-prl-98} considered a single movable
quantum dot, while Armour and MacKinnon \cite{arm-prb-02}
introduced a model of a three-dot array whose central dot is
movable, see Fig.\ref{setup}. Both of these systems are of
intrinsic interest and may find applications in real systems. The
quantum system is assumed to be in the strong Coulomb blockade
regime in which only none or one extra (spinless) electron in the
whole system are allowed. Thus, in the one-dot case the electronic
states are $|0\rangle$ and $|1\rangle$, while in the triple-dot
case we have $|0\rangle, |L\rangle, |C\rangle,$ and $|R\rangle$.
The quantum system is coupled to two leads with a high bias
applied between them. The bias is smaller than the charging energy
for addition or removal of other electrons but otherwise it is the
largest energy scale in the model.

The moving dot interacts with its surroundings and its dissipative
dynamics is described by the interaction with a generic heat bath.
The Hamiltonian has the form
\begin{equation}\label{Hamiltonian}
    \hat{H} = \hat{H}_{\rm osc} + \hat{H}_{\rm el}+\hat{H}_{\rm el-leads}(\hat{x})
              + \hat{H}_{\rm leads}
            + \hat{H}_{\rm bath}+ \hat{H}_{\rm osc-bath}
\end{equation}
where
\begin{equation}
       \hat{H}_{\rm osc}=\frac{\hat{p}^2}{2m}+\frac{m\om_0^2\hat{x}^2}{2}
       \tag{1a}
\end{equation}
describes the mechanical center-of-mass motion of the central dot
as a one-dimensional harmonic oscillator with mass $m$ and
frequency $\om_0$. We emphasize the importance of the nonlinear
dependence of  $\hat{H}_{\rm el-leads}$  on the oscillator degree
of freedom; the $\hat{x}$-dependence is often exponential which
lies at the heart of the shuttling instability. We refer to the
literature for explicit expressions for the other terms appearing
in Eq.(\ref{Hamiltonian}) \cite{gor-prl-98,arm-prb-02,nov-prl-03}.
The leads are held at different electrochemical potentials
$\mu_{L,R}$ whose difference gives the bias across the array (see
Fig. (\ref{setup})). We assume that the tunneling densities of
states are independent of energy. This is necessary for the  {\em
first Markov approximation},\cite{gardiner} used later on, to
hold. Further, as already implied above, we assume
$\mu_L\to\infty,\,\mu_R\to -\infty$. These assumptions are
necessary for the derivation of the Markovian dynamics of the
movable system.
Finally, the heat bath consisting of an infinite set of harmonic
oscillators linearly coupled to the mechanical degree of freedom
is described with the Caldeira-Leggett model\cite{weiss} in its
Ohmic form.

\subsection{Generalized Master Equation}
\label{GMEsection}

For the description of the model we use the language of quantum
dissipative systems.\cite{weiss} As the ``system" (or ``device")
we take the electronic states plus the one-dimensional oscillator
describing the center-of-mass motion of the central dot. The
electronic leads and the heat bath interacting with the mechanical
degree of freedom constitute the reservoirs. The task is  to
integrate out the degrees of freedom of the reservoirs to end up
with an equation of motion for the system density operator. The
derivation proceeds in two steps first integrating out the leads
and then the heat bath in the weak coupling limit to get the
desired GME for the system density operator.  We remark that the
assumed additivity of the two heat baths should be proven, however
we are not aware of any such proof.

As a specific example of an end result of this procedure, we give
the GME governing the system density operator in the one-dot case,
written here in a Liouvillean form:
\begin{align}
   \dot{\rho}(t) &= \mathcal{L}\rho(t) = (\mathcal{L}_{\rm coh} + \mathcal{L}_{\rm driv}
   + \mathcal{L}_{\rm damp})\rho(t) \label{supermatrix}\\
\intertext{where the various {\it superoperators} $\mathcal{L}_i$,
$i$= coh, driv, damp, are defined as}
   \mathcal{L}_{\rm coh}\rho &= \frac{1}{i\hbar}[H_{\rm osc}
   + \eps_0 c^{\dag}_0 c_0- eE x c^{\dag}_0 c_0,\rho] \ ,\\
   \mathcal{L}_{\rm driv}\rho &= -\frac{\Gamma_L}{2}
   \bigl(c_0 c^{\dag}_0 e^{-\frac{2x}{\lambda}}\rho -2 c^{\dag}_0 e^{-\frac{x}{\lambda}}
   \rho e^{-\frac{x}{\lambda}} c_0 + \rho e^{-\frac{2x}{\lambda}}c_0
   c^{\dag}_0\bigr)  \notag \\
   &-\frac{\Gamma_R}{2}
   \bigl(c^{\dag}_0 c_0 e^{\frac{2x}{\lambda}}\rho -2 c_0 e^{\frac{x}{\lambda}}
   \rho e^{\frac{x}{\lambda}}c^{\dag}_0 + \rho e^{\frac{2x}{\lambda}}c^{\dag}_0
   c_0\bigr) \ , \\
   \mathcal{L}_{\rm damp}\rho &= -\frac{i\gamma}{2\hbar}[x,\{p,\rho\}] -
   \frac{\gamma m \omega}{\hbar}(\bar{N}+1/2)[x,[x,\rho]] \ .\label{TI}
\end{align}
The physical meaning of certain of the terms is of particular
interest.  The term proportional to $E$ is due to the
electrostatic force on an occupied (i.e., charged) dot.  The
exponential dependence of the tunneling terms is clearly visible.
The damping of the oscillator cannot be described entirely
satisfactorily in the Markovian limit: one cannot simultaneously
achieve translational invariance, positivity of the density
matrix, and relaxation towards canonical equilibrium.  The form
given above does not satisfy strict positivity; the quantitative
consequences of this shortcoming turned out to be negligible
\cite{nov-prl-03}.  To proceed further, one considers the
electronic diagonal elements $\rho_{00}(t)=\langle
0\,|\rho(t)|0\rangle$ and $\rho_{11}(t)=\langle
1\,|\rho(t)|1\rangle$, where $|1\rangle=c_0^{\dag}|0\rangle$.
These objects are still full density matrices in the phonon space
and satisfy
\begin{equation}\label{GME}
\begin{split}
    \dot{\rho}_{00}(t) &= \frac{1}{i\hbar} [H_{\rm osc},\rho_{00}(t)]
    - \frac{\Gamma_L}{2}(e^{-\frac{2x}{\lambda}}\rho_{00}(t)
    + \rho_{00}(t)e^{-\frac{2x}{\lambda}})
    + \Gamma_R e^{\frac{x}{\lambda}}\rho_{11}(t)e^{\frac{x}{\lambda}}
    + \mathcal{L}_{\rm damp}\,\rho_{00}(t)\ , \\
    \dot{\rho}_{11}(t) &= \frac{1}{i\hbar}[H_{\rm osc}-eEx,\rho_{11}(t)]
    + \Gamma_L e^{-\frac{x}{\lambda}}\rho_{00}(t)e^{-\frac{x}{\lambda}}
    - \frac{\Gamma_R}{2}(e^{\frac{2x}{\lambda}}\rho_{11}(t)
    + \rho_{11}(t)e^{\frac{2x}{\lambda}})
    + \mathcal{L}_{\rm damp}\,\rho_{11}(t).
\end{split}
\end{equation}

\section{Curent and noise}
\subsection{Analytical development}
The current through the system is given by
\begin{equation}\label{current}
    I^{\rm stat}=e \Gamma_L{\rm Tr_{osc}}(e^{-\frac{2x}{\lambda}}\rho^{\rm
    stat}_{00})
    = e \Gamma_R{\rm Tr_{osc}}(e^{\frac{2x}{\lambda}}\rho^{\rm
    stat}_{11}) \ .
\end{equation}
The trace is carried out over the oscillator basis and $\rho^{\rm
stat}_{nn} = \lim_{t\to\infty}\rho_{nn}(t)$ .  To find the
stationary solution, one needs to truncate the oscillator basis at
some suitably large value $N$ (in practice, $N$ may reach 100
before convergence is achieved); since the resulting linear system
has dimensions $2N^2\times 2N^2$ (single-dot), or $10N^2\times
10N^2$ (triple-dot), this becomes a critical issue. The method of
solving this vast numerical task is commented on below.

Let us next consider the calculation of noise.  In quantum optics
one often resorts to a result known as the Quantum Regression
Theorem; this allows the calculation of any multi-time correlation
function of {\it system operators}. Unfortunately, in NEMS this
theorem can be applied only under very restricted circumstances:
it {\it can} be used to calculate the noise (which is essentially
the average of a product of current operators) {\it within} the
three-dot device\cite{fli-prb-04} (because the current operators
$\hat{I}_{LC}$ and $\hat{I}_{RC}$ operate within the ``system"),
while it is unapplicable for the single-dot case (and many other
NEMS systems as well) because there the current operators involve
operators belonging both to the baths (i.e., the electronic leads)
and the system. A more general method is thus called for.

In order to compute the noise spectrum, we follow the ideas of
Gurvitz and Prager \cite{gur-prb-96}, and introduce
number-resolved density-matrices $\rho_{ii}^{(n)}$, where $n$ is
the number of electrons tunnelled into the right lead by time $t$.
Obviously, $\rho_{ii}(t)=\sum_n \rho_{ii}^{(n)}(t)$. The
$\rho_{ii}^{(n)}$ obey
\begin{equation}\label{GMEn}
\begin{split}
    \dot{\rho}_{00}^{(n)}(t) &= \frac{1}{i\hbar} [H_{\rm osc},\rho_{00}^{(n)}(t)]
    + \mathcal{L}_{\rm damp}\,\rho_{00}^{(n)}(t)\\
    &-
    \frac{\Gamma_L}{2}\{e^{-\frac{2x}{\lambda}},\rho_{00}^{(n)}(t)\}
    + \Gamma_R e^{\frac{x}{\lambda}}\rho_{11}^{(n-1)}(t)e^{\frac{x}{\lambda}}
    \ , \\
    \dot{\rho}_{11}^{(n)}(t) &= \frac{1}{i\hbar}[H_{\rm osc}-eEx,\rho_{11}^{(n)}(t)]
    + \mathcal{L}_{\rm damp}\,\rho_{11}^{(n)}(t)\\
    &-
    \frac{\Gamma_R}{2}\{e^{\frac{2x}{\lambda}},\rho_{11}^{(n)}(t)\}
    + \Gamma_L e^{-\frac{x}{\lambda}}\rho_{00}^{(n)}(t)e^{-\frac{x}{\lambda}}
    \ ,
\end{split}
\end{equation}
with $\rho_{11}^{(-1)}(t)\equiv 0$. The mean current and the
zero-frequency shot noise spectrum are given by \cite{ela-pla-02}
\begin{align}
    &I = e \frac{d}{dt}\sum_nn P_n(t)\Big|_{t\to\infty}
    = e \sum_n n \dot{P}_n(t)\Big|_{t\to\infty},\label{current2}\\
    &S(0) = 2e^2\frac{d}{dt}\bigg[\sum_nn^2P_n(t)
    -\Big(\sum_nnP_n(t)\Big)^2\bigg]\bigg|_{t\to\infty}
    \label{macdonald},
\end{align}
where
$P_n(t)=\mathrm{Tr_{osc}}[\rho_{00}^{(n)}(t)+\rho_{11}^{(n)}(t)]$
are the probabilities of finding $n$ electrons in the right lead
by time $t$, i.e. precisely the objects needed for the FCS
discussed below. We find $I=\sum_n n \dot{P}_n(t)
=\Gamma_R\mathrm{Tr_{osc}}\big(e^{\tfrac{2x}{\lambda}}\rho_{11}(t)\big)$,
i.e. one recovers the stationary current found above. In a similar
fashion, $\sum_n n^2 \dot{P}_n(t)
=\Gamma_R\mathrm{Tr_{osc}}\big[e^{\tfrac{2x}{\lambda}}\big(2\sum_n
n \rho_{11}^{(n)}(t) + \rho_{11}(t)\big)\big]$, whose large-time
asymptotics determines the shot noise according to
(\ref{macdonald}).  We have developed a generating function
technique in Ref. \onlinecite{nov-prl-04} to extract this
large-time limit.  Here we skip the technical details; the upshot
is that the zero-frequency noise, and thereby the Fano factor
$F\equiv S(0)/2eI$, can be expressed in terms of the pseudoinverse
of the Liouvillean (and the static limit of the density matrix
known already from the current calculation):
\begin{equation}\label{Fano}
    F = 1 - \frac{2e\Gamma_R}{I} \mathrm{Tr_{osc}}\left(e^{\frac{2x}{\lambda}}
    \left[\mathcal{Q}\mathcal{L}^{-1}\mathcal{Q}\begin{pmatrix}
    \Gamma_R e^{\frac{x}{\lambda}}\rho_{11}^{\rm
    stat}e^{\frac{x}{\lambda}}\\
    0 \end{pmatrix}\right]_{11}\right)\ .
\end{equation}
Here $\mathcal{Q}$ is a projection operator that projects away
from the stationary state, for which the Liouvillean has the
eigenvalue zero.   The crucial point is that the pseudoinverse
$\mathcal{R}$ of the Liouvillean, defined as
$\mathcal{Q}\mathcal{L}^{-1}\mathcal{Q}\equiv\mathcal{R}$ is
tractable by similar numerical methods as used in the evaluation
of the current.  In our discussion of the full counting statistics
given below, we analyze a toy-model to illustrate some properties
of the pseudoinverse $\mathcal{R}$.

\subsection{Comment on numerics; some results}

As mentioned above, the superoperator structure of the Liouville
equation leads to large matrices of the order of $N^2\times N^2$,
where $N$ is the number of low-energy states kept in the
calculation.  A further complication arises from the fact that the
stationary limit corresponds to the zero eigenvalue of the
Liouvillean, $\mathcal{L}\rho=0$, forcing one to deal with
singular matrices.  The problems with the memory size can be
circumvented by using iterative methods in which only
$\mathcal{L}A$ for a given $A$ is needed ($N\times N$ numbers),
and one avoids the storage of the full $\mathcal{L}$ ($N^2 \times
N^2$ numbers).  Using an iterative method raises the questions of
convergence, and the speed of convergence.  We found that the
so-called Arnoldi iteration technique was sufficient for our
purposes (see Appendix A in Ref. \onlinecite{fli-prb-04}),
provided that one uses a suitable preconditioning.   While there
exists a substantial empirical body of knowledge of how to carry
out the preconditioning, we are not aware of a complete algorithm.
As an example, when solving for the pseudoinverse $\mathcal{R}$,
we used the inverse of the ``Sylvester part" $\mathcal{L}_0$ of
$\mathcal{L}$ as preconditioner\cite{andrea}; finding the inverse
of $\mathcal{L}_0$ is a relatively fast procedure. When
calculating the noise (which amounts to solving an equation of the
type $\mathcal{L}\mathbf{x}=\mathbf{b}$, where the vector
$\mathbf{b}$ belongs in the range of $\mathcal{L}$), we used the
generalized minimum residual method (GMRes)\cite{christian}.
Again, appropriate preconditioning was crucial.

\begin{figure}
 \includegraphics[width=\textwidth]{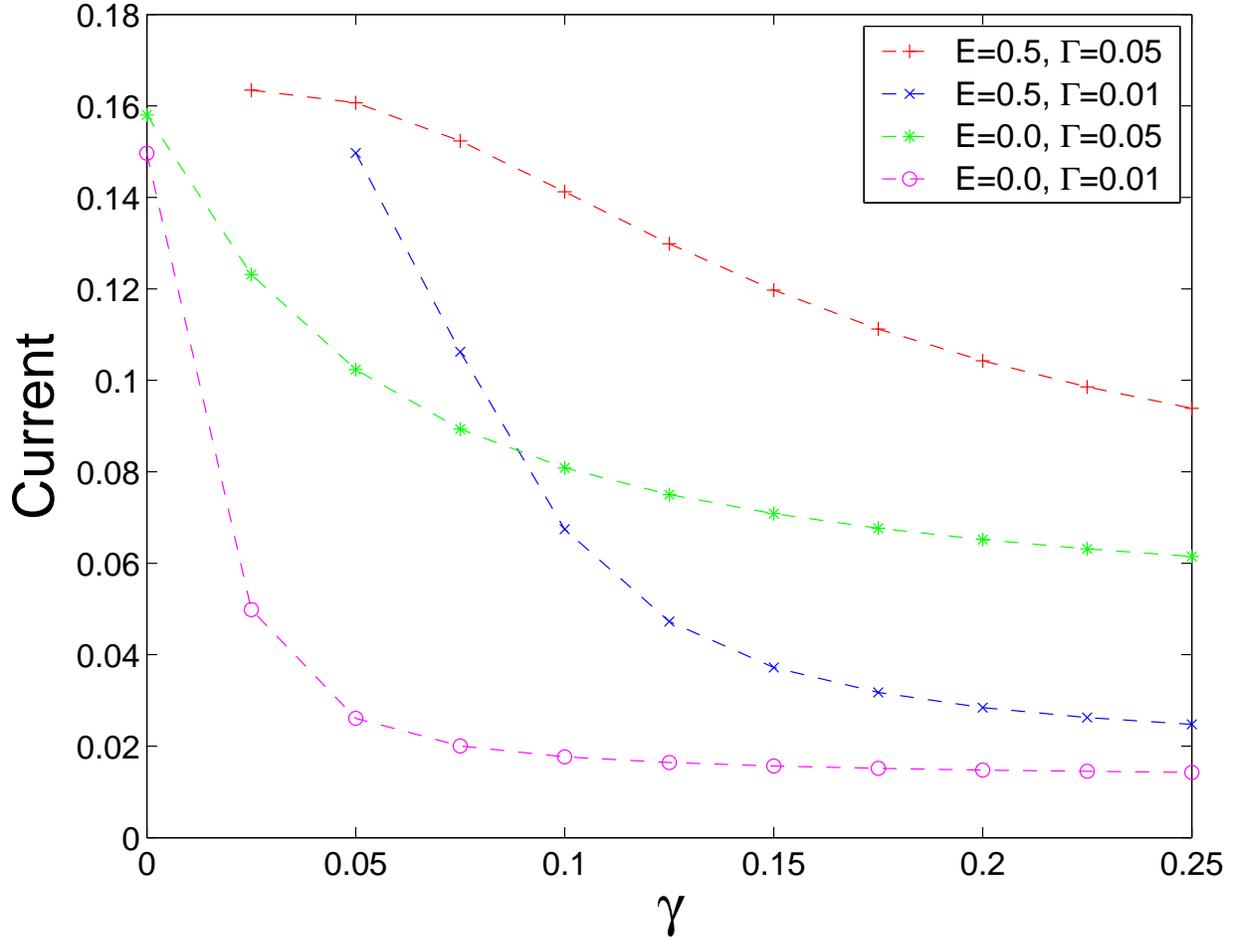}
  \caption{$I-\gamma$ curve. The damping dependence of the stationary
current through the single-dot shuttle for different transfer
rates and electric fields. Their values are
$d=0.5x_0,\Gamma=0.05\hbar\om$ (pluses; corresponds to
Fig.~\ref{fig1prl03}), $d=0.5x_0,\Gamma=0.01\hbar\om$ (circles),
$d=0.0,\Gamma=0.05\hbar\om$ (asterisks), $d=0.0,\Gamma=0.01\hbar$
(crosses). Other parameters are $\lambda=x_0,T=0$. The current is
in units of $e\om$ while $\gamma$ in $\hbar\om$. (Reproduced from
Ref.\onlinecite{nov-prl-03})}\label{fig2prl03}
\end{figure}

Figure \ref{fig2prl03} shows a set of current vs. damping curves
for the single-dot shuttle.  We draw attention to the following
features.  As damping is decreased, the current increases,
approaching asymptotically the value $I=1/2\pi\simeq 0.16$, i.e.
one electron is transferred per cycle.  In other words, this value
of the current indicates that the shuttling transition has taken
place, even in the quantum regime.  The transition is not as sharp
as found in the classical case\cite{gor-prl-98}.  For large values
of damping the current is much smaller, and scales with the
tunneling rate $\Gamma$.  Very interestingly, we see a sharp
increase in the current even without electric field, $E=0$.
Classically this does not happen, and therefore we interpret this
cross-over of being due to quantum shot noise.

\begin{figure}
  \includegraphics[width=\textwidth]{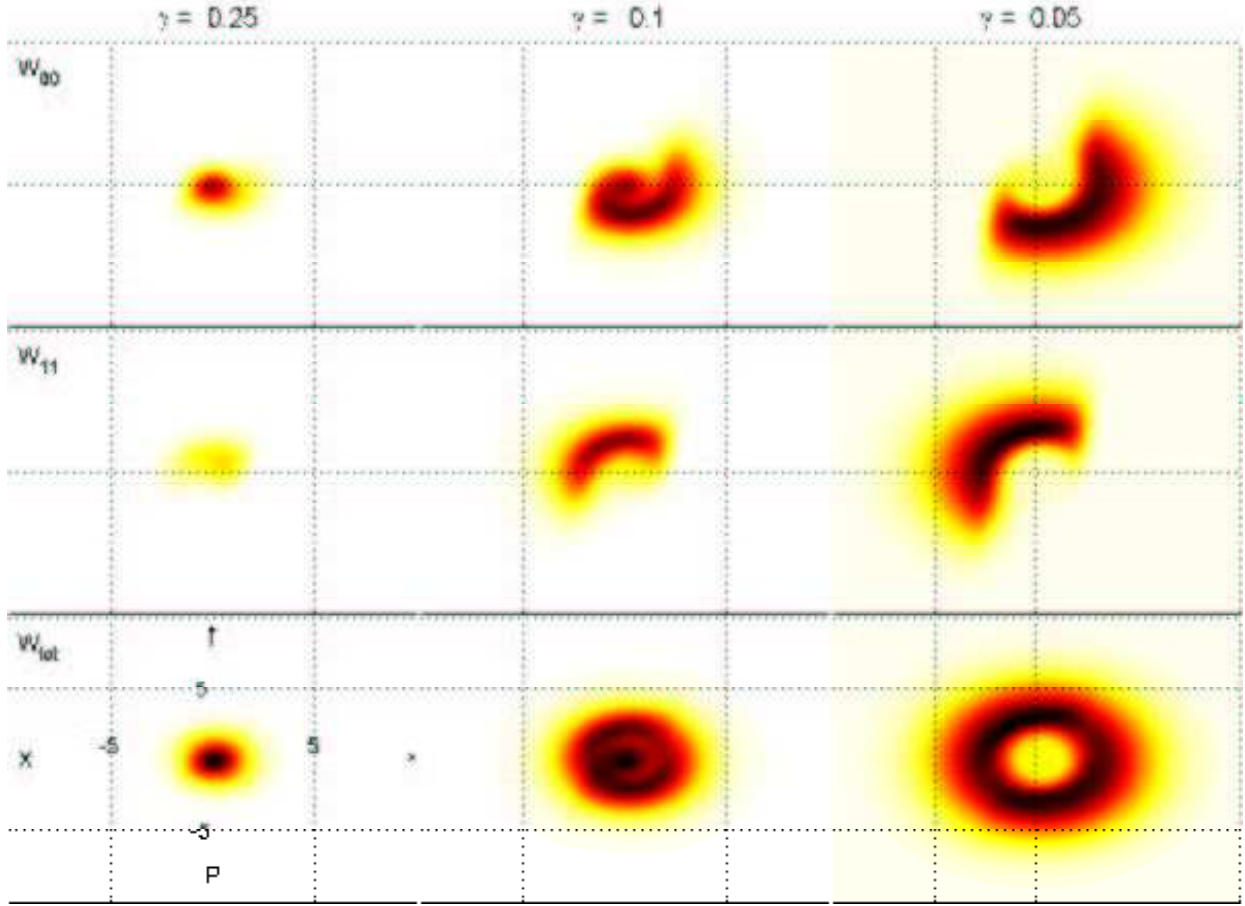}
  \caption{Phase space picture of the tunnelling-to-shuttling crossover.
  The respective rows show the Wigner distribution functions
  for the discharged ($W_{00}$), charged ($W_{11}$), and both ($W_{\rm tot}$)
  states of the oscillator in the phase space (horizontal axis --
  coordinate in units of $x_0=\sqrt{\hbar/m\om}$,  vertical axis -- momentum in $\hbar/x_0$).
  The values of the parameters are: $\lambda=x_0,T=0,d=0.5x_0,\Gamma=0.05\hbar\om$.
  The values of $\gamma$ are in units of $\hbar\om$. The Wigner functions are
  normalized within each column. (Reproduced from Ref. \onlinecite{nov-prl-03})}
  \label{fig1prl03}
\end{figure}

\begin{figure}
  \includegraphics[height=\textwidth,angle=-90]{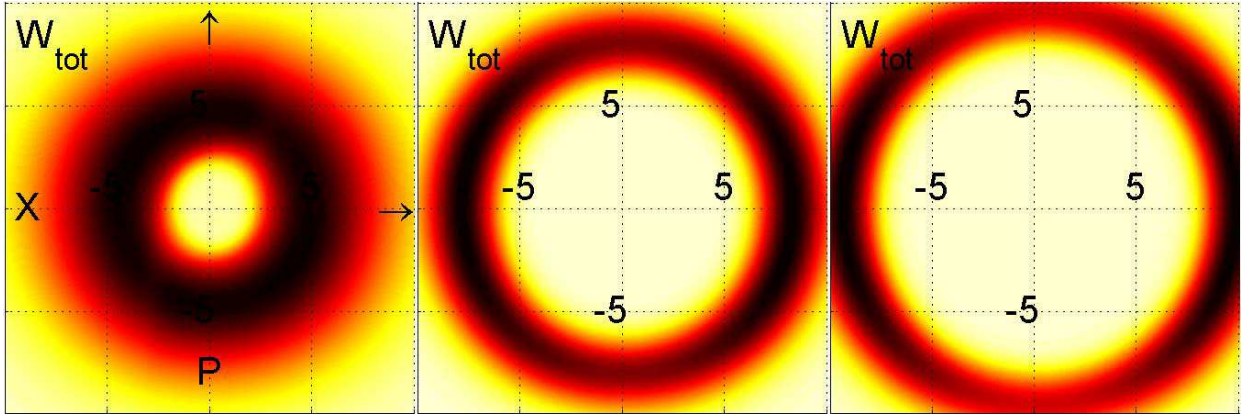}
  \caption{Transition from quantum limit to classical limit.  The ``quantum
  thickness"
  of the Wigner ring shrinks, and the radius increases, indicating larger maximal
  velocity, and oscillation amplitude.}
  \label{wignerquantclass}
\end{figure}

\begin{figure}
 \includegraphics[width=\textwidth]{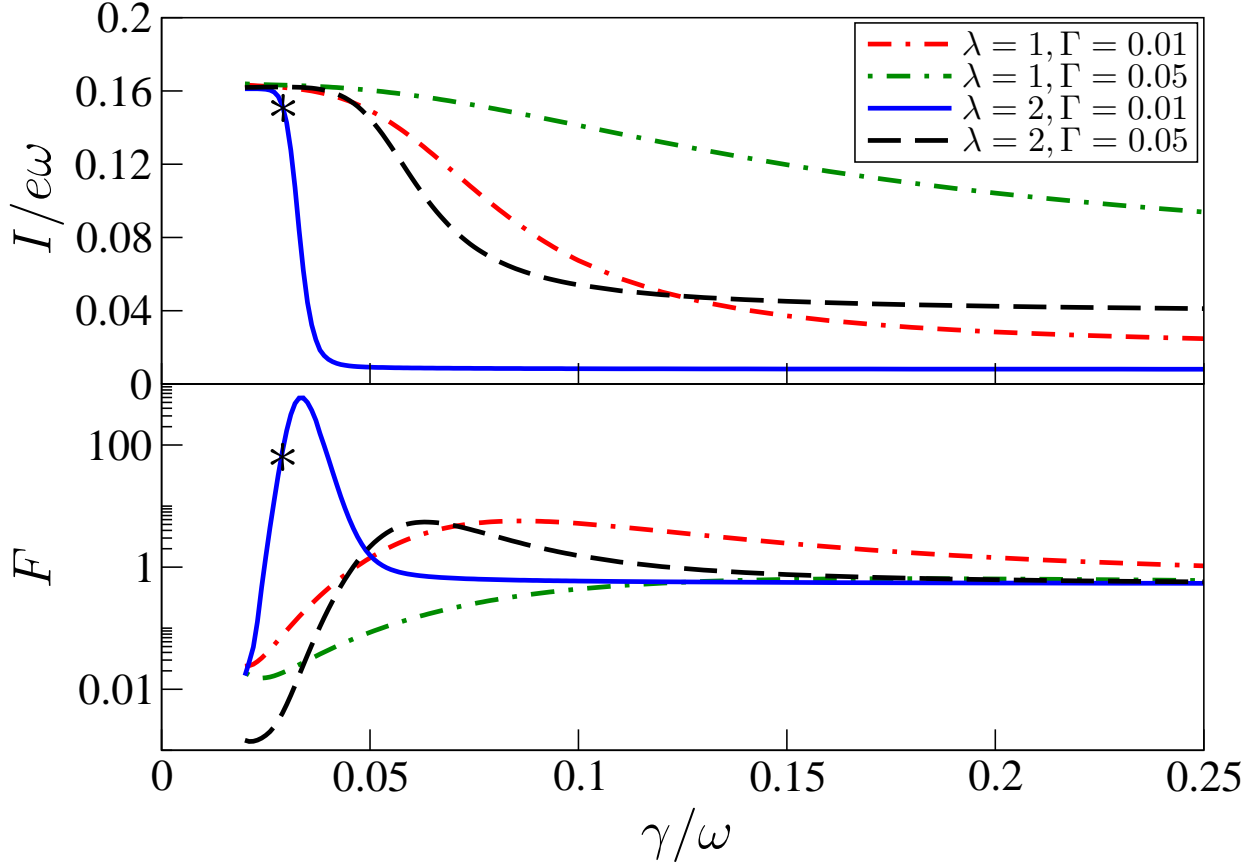}
  \caption{Current $I$ and Fano factor $F$ vs.\ damping $\gamma$.
The $\gamma$-dependence of $I$ (upper panel) and $F$ on log scale
(lower panel) for different transfer rates $\Gamma$ and tunnelling
lengths $\lambda$. The parameters are $\lambda=x_0,\Gamma=0.01\om$
(full); $\lambda=x_0,\Gamma=0.05\om$ (long dashes);
$\lambda=2x_0,\Gamma=0.01\om$ (short dashes);
$\lambda=2x_0,\Gamma=0.05\om$ (dots) with $x_0=\sqrt{\hbar/m\om}$.
Other parameters are $eE/m\om^2=0.5x_0,\,T=0$. The current is in
units of $e\om$ while $\gamma$ in units of $\om$. The asterisk
defines the parameters of Wigner distributions in
Fig.~\ref{fig2-prl04}. (Reproduced from Ref.
\onlinecite{nov-prl-04}.)} \label{fano_fig}
\end{figure}

We have in several occasions promoted the use of Wigner functions
as an interpretative tool for the numerical results obtained for
the stationary density matrix.  The Wigner representation of the
GME has also turned out to be a useful starting point for further
analytic work\cite{fed-prl-04,fed-preprint-04,andrea}. These
phase-space representations have a simple form in the classical
limit: the Wigner representation of a regularly moving harmonic
oscillator is an ellipse.  On the other hand, irregular motion
under the influence of external noise gives rise to a Gaussian
probability distribution centered at the origin. The {\it charge
resolved} Wigner functions ($n=0$ corresponds to an empty dot,
while $n=1$ represents the occupied dot)  are defined as

\begin{equation}
    W_{nn}(X,P) =
    \int_{-\infty}^{\infty}\frac{dy}{2\pi\hbar}\,\bigl\langle
    X-\frac{y}{2}\,|\rho_{nn}^{\rm stat}|X+\frac{y}{2}\bigr\rangle\,
    \exp\bigl(i\frac{Py}{\hbar}\bigr),
\end{equation}
and some representative results are given in Fig. \ref{fig1prl03}.
As expected, as damping is decreased, the fuzzy central spot
evolves into a ring.  The finite thickness of the ring is due to
thermal noise, the randomness of the charge transport processes,
and the position--momentum uncertainty. The earmark of shuttling
are the asymmetric, banana-shaped areas observed for weakest
damping: here one observes a strong correlation between the
occupancy of the dot, and the position and momentum.  Thus, there
is a large probability to have an occupied dot with a negative
position coordinate and positive velocity (the dot has been filled
in the neighborhood of the left contact), while there is a large
probability of having an empty dot at positive $x$-values and
negative velocity, i.e. on the return journey from the right
contact. -- As the system approaches the classical limit, the
thickness of the Wigner ring shrinks; this is illustrated in Fig.
\ref{wignerquantclass}. As mentioned above, we expect that the
noise of the quantum shuttle yields additional information about
the nature of the charge transfer process. Figure \ref{fano_fig}
shows some of our numerical results.  In the top panel we see a
relatively sharp cross-over if the tunneling length is larger than
the length scale $x_0=\sqrt{\hbar/(m\omega)}$, i.e. as one
aproaches the classical limit.  In the bottom panel we show the
Fano factor for the same parameters as in the top panel.  One
should note that a logarithmic scale is used, thus one observes
(i) a huge enhancement at the tunneling cross-over, in particular
for parameters approaching the semiclassical limit, and (ii) tiny
Fano factors below the shuttling cross-over.  Thus, even well in
the quantum regime shuttling is a highly ordered charge transfer
process.

\begin{figure}
  \includegraphics[width=\textwidth]{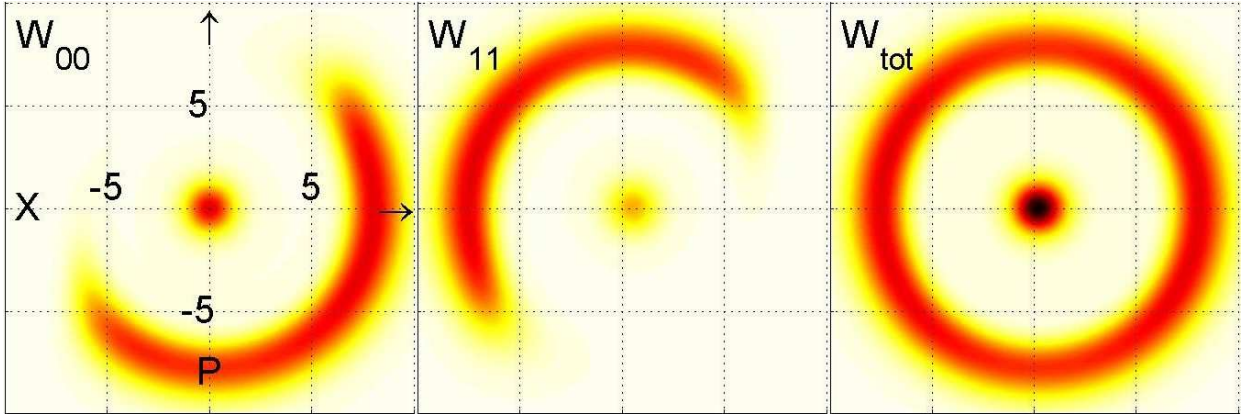}
  \caption{Phase space picture of the shuttle around the transition
  where the shuttling and tunnelling regimes coexist.
  The respective rows show the Wigner distribution functions
  for the discharged ($W_{00}$), charged ($W_{11}$), and both ($W_{\rm tot}=W_{00}+W_{11}$)
  states of the oscillator in the phase space (horizontal axis --
  coordinate in units of $x_0=\sqrt{\hbar/m\om}$,  vertical axis -- momentum in $\hbar/x_0$).
  The values of the parameters are: $\lambda=2x_0,eE/m\om^2=0.5x_0,\gamma=0.029\om,
  \Gamma=0.01\om,T=0$.}
  \label{fig2-prl04}
\end{figure}

Figure \ref{fig2-prl04} shows the phase-space plot for the
parameter values indicated by the asterisk in Fig. \ref{fano_fig}.
The interpretation is very suggestive: for these parameters
tunneling and shuttling {\it co-exist}.  Our numerics thus confirm
the suggestion put forward in the analytical study of Ref.
\onlinecite{fed-prl-04}.  The large value of the Fano factor can
be understood as a slow switching process between the two possible
current channels (tunneling and shuttling).  In the next section
we elaborate this point further.
\section{Full counting statistics}
\subsection{Calculational procedure}
\begin{figure}
  \includegraphics[width=\textwidth]{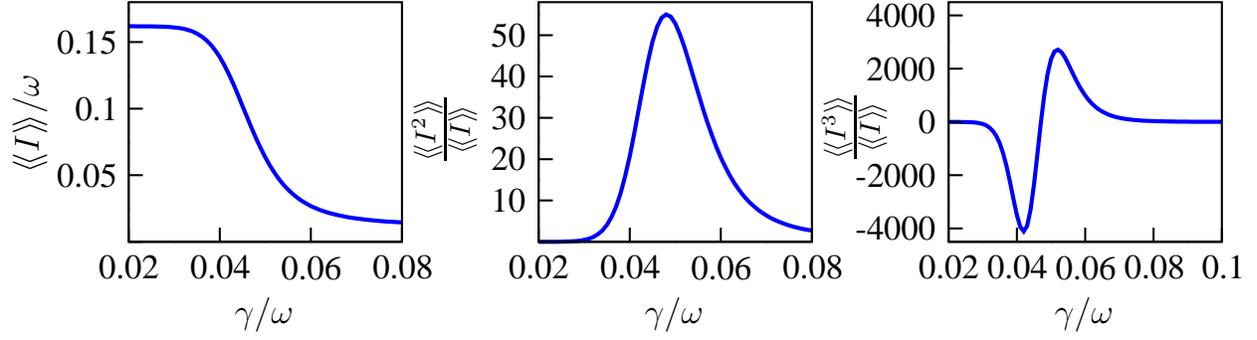}
  \caption{The first three cumulants for the one-dot shuttle as a function of
  the damping $\gamma$.  The parameters are $\lambda=1.5 x_0$, $d=eE/m\omega^2=0.5 x_0$. }
  \label{thirdcumulant}
\end{figure}

We begin by introducing some notation.  We recall that the
Liouvillean, which is a non-hermitian operator, has a single
eigenvalue equal to zero with $\hat{\rho}^{\rm{stat}}$ being the
corresponding (normalized and unique) right eigenvector, we denote
this eigenvector by $|0\rrangle$. The corresponding left
eigenvector is the identity operator $\hat{1}$ which we denote by
$\llangle\tilde{0}|$, and we have
$\llangle\tilde{0}|0\rrangle\equiv\rm{Tr}(\hat{1}^{\dagger}\hat{\rho}^{\rm{
stat}})=1$. The pair of eigenvectors allows us to define the
complementary projectors $\mathcal{P}\equiv|0\rlangle\tilde{0}|$
and $\mathcal{Q}\equiv1-\mathcal{P}$ obeying the relations
$\mathcal{PL}=\mathcal{LP}=0$ and $\mathcal{QLQ}=\mathcal{L}$. We
will also need the pseudoinverse of the Liouvillean
$\mathcal{R}\equiv\mathcal{Q}\mathcal{L}^{-1}\mathcal{Q}$, which
is well-defined, since the inversion is performed only in the
subspace spanned by $\mathcal{Q}$, where $\mathcal{L}$ is regular.
Rather than attempting to calculate the complete probability
distribution $P_n(t)=\rm{Tr}\left[\hat{\rho}^{(n)}(t)\right]$
directly, it turns out to be easier to evaluate the cumulant
generating function $S(t,\chi)$,
\begin{equation}
e^{S(t,\chi)}=\sum_{n=-\infty}^{\infty}P_n(t)e^{in\chi}.
\end{equation}
From $S(t,\chi)$ we find the $m$'th cumulant of the charge
distribution by taking the $m$-th derivative with respect to the
counting field $\chi$ at $\chi=0$:
\begin{equation} \llangle
n^m\rrangle(t)=\frac{\partial^mS}{\partial(i\chi)^m}|_{\chi=0},
\end{equation}
and from the knowledge of all cumulants we can reconstruct
$P_n(t)$. The cumulants of the current in the stationary limit
$t\to\infty$ are given by the time derivative of the charge
cumulants:
\begin{equation}
\llangle I^m\rrangle=\frac{d}{dt}\llangle n^m\rrangle(t)\big|_{\,
t\to\infty}.
\end{equation}
The first two current cumulants give the average current running
through the system and the zero-frequency current noise,
respectively. We  have recently developed a systematic
perturbation theory to calculate the higher
cumulants\cite{fli-epl-04}, and we just quote  the final results:

\begin{eqnarray}
\llangle I\rrangle &=& \llangle\tilde{0}|\mathcal{I}\,|0\rrangle,
\nonumber\\
\llangle I^2\rrangle &=&
\llangle\tilde{0}|\mathcal{J}|0\rrangle-2\llangle\tilde{0}|\mathcal{I}\mathcal{R}\mathcal{I}\,|0\rrangle,
\nonumber\\
\llangle I^3\rrangle &=& \llangle\tilde{0}|\mathcal{I}\,|0\rrangle
-3\llangle\tilde{0}|\mathcal{I}\mathcal{R}\mathcal{J}+\mathcal{J}\mathcal{R}\mathcal{I}\,|0\rrangle
-6\llangle\tilde{0}|\mathcal{I}\mathcal{R}(\mathcal{R}\mathcal{I}\mathcal{P}-
\mathcal{I}\mathcal{R})\mathcal{I}\,|0\rrangle,
\label{eq_thirdcumulant}
\end{eqnarray}
where
\begin{eqnarray}
\mathcal{I}&=& \mathcal{I}^{+}-\mathcal{I}^{-}\\
 \mathcal{J}&=&
\mathcal{I}^{+}+\mathcal{I}^{-}.
\end{eqnarray}
The explicit expressions for the current superoperators
$\mathcal{I}^\pm$ depend on the problem at hand; two examples are
given in Ref. \onlinecite{fli-epl-04} (see also the analytic toy
model discussed below).  We emphasize the utility of these
formulae: they imply that once the hard numerical problems with
the stationary case are solved, the higher cumulants are
essentially immediately accessible. Using the systematic
perturbation theory developed in Ref. \onlinecite{fli-epl-04}
algebraic formulas for the fourth, fifth etc. cumulant can be
generated with computer routines performing symbolic
manipulations.  In Fig. \ref{thirdcumulant} we show numerical
results for the first three cumlants for the single-dot shuttle.

\subsection{Toy model}

The general formulation for calculation of higher cumulants given
above is very practical also in analytic calculations.  Here we
illustrate it by considering a toy model, for which the current
and noise are very well-known, while the third cumulant cannot be
considered as a standard result.  Specifically, we consider a two
level system, whose occupation probabilities obey the Master
equation (${\mathbf p}=(p_0\quad p_1)^{\rm T}$)
\begin{equation}
\dot{\mathbf p}={\mathcal L}{\mathbf p}=\begin{pmatrix} -\Gamma_L&\Gamma_R\\
\Gamma_L & -\Gamma_R.
\end{pmatrix}{\mathbf p}\label{2-level}
\end{equation}
The relevance of this model to the shuttling dynamics is based on
the fact that in the co-existence regime there are two ``states":
the incoherent tunneling current channel (reflected by the central
spot in the Wigner function description), and the orderly
shuttling channel (reflected) by the ring, and that the system can
be visualized as fluctuating between these two states.  The bare
two-level system, as discussed in this section, needs a slight
modification to take into account that in the coexistence regime
one deals with {\it current channels} and not states, and further
discussion can be found in Ref. \onlinecite{fli-epl-04}, see also
Ref. \onlinecite{jor-preprint-04} for a related model.  The
Liouvillean of Eq. (\ref{2-level}) has the following right and
left null vectors:
\begin{equation}
|0\rrangle=\frac{1}{\Gamma_L+\Gamma_R}\begin{pmatrix}\Gamma_R \\
\Gamma_L\end{pmatrix},\quad \llangle{\tilde 0}|=\begin{pmatrix}1 &
1\end{pmatrix},
\end{equation}
which satisfy $\LL |0\rrangle=0=\llangle{\tilde 0} |\LL$ and
$\llangle{\tilde 0}|0\rrangle=1$, as is readily verified.  We can
immediately form the projectors $\PP$ and $\QQ$:
\begin{equation}
\PP=|0\rrangle \! \llangle{\tilde
0}|=\frac{1}{\Gamma_R+\Gamma_L}\begin{pmatrix} \Gamma_R & \Gamma_R
\\ \Gamma_L & \Gamma_R \end{pmatrix},\quad
\QQ=1-\PP=\frac{1}{\Gamma_R+\Gamma_L}
\begin{pmatrix} \Gamma_L & -\Gamma_R \\ -\Gamma_L & \Gamma_R
\end{pmatrix}.
\end{equation}
We also record the current superoperator:
\begin{equation}
{\mathcal I}_{0R}=\begin{pmatrix} 0 & \Gamma_R \\ 0 & 0
\end{pmatrix}.
\end{equation}
Next, we  need the eigenvector $|\nu\rrangle$ of $\LL$ which does
not belong to the null space (here, the situation is quite simple
because the full problem is 2-dimensional with a 1-dimensional
null space), and the associated eigenvalue $\lambda_\nu$.  We
readily find
\begin{equation}
\lambda_\nu=-(\Gamma_R+\Gamma_L);\quad
|\nu\rrangle=\begin{pmatrix}1 \\ 1 \end{pmatrix}; \quad
\llangle{\tilde
\nu}|=\frac{1}{\Gamma_R+\Gamma_L}\begin{pmatrix}\Gamma_L &
\Gamma_R \end{pmatrix}, \end{equation} and can thus compute the
pseudoinverse $\mathcal R$:
\begin{equation}
{\mathcal R}=\QQ {\LL}^{-1} \QQ =
\frac{1}{\lambda_\nu}|\nu\rrangle \llangle {\tilde
\nu}|=-\frac{1}{(\Gamma_R+\Gamma_L)^2}\begin{pmatrix} \Gamma_L &
-\Gamma_R \\ -\Gamma_L & \Gamma_R
\end{pmatrix}=\frac{1}{\lambda_\nu} \QQ.
\end{equation}
With these results at hand, the evaluation of various cumulants is
reduced to simple matrix multiplications, and we quote the results
\begin{eqnarray}
\llangle {\mathcal I} \rrangle &=&
\frac{\Gamma_L\Gamma_R}{\Gamma_L+\Gamma_R}\\
\llangle {\mathcal I}^2 \rrangle &=&
\frac{\Gamma_R^2+\Gamma_L^2}{(\Gamma_R+\Gamma_L)^2}\,\llangle
{\mathcal I}\rrangle\\ \llangle {\mathcal I}^3 \rrangle &=&
\frac{\Gamma_R^4-2\Gamma_R^3\Gamma_L+6\Gamma_R^2\Gamma_L^2-
2\Gamma_R\Gamma_L^3+\Gamma_L^4}{(\Gamma_R+\Gamma_L)^2}\,\llangle
{\mathcal I}\rrangle.
\end{eqnarray}
The first two results are well-known from the literature (see,
e.g., Ref. \onlinecite{bla-phr-00}), however the third cumulant
may be new.  We believe that the above method of calculation can
be very useful in extending several previous results, such as
those obtained in by Kie{\ss}lich et al.\cite{Kie-prb-03}.
\subsection{Experimental status}
To summarize the experimental status very shortly, we do not
believe that the shuttling transition has been observed so far.
Having stated this negative conclusion, we hasten to point out
that we think that the experimental situation is very promising,
and that a crucial experiment could be just behind the corner.

Two key experiments have been reported, which contain several
important ingredients.  In the experiment of Park {\it et
al.}\cite{par-nat-00} a ${\rm C}_{60}$ molecule was placed in a
break junction, and the current-voltage characteristics showed
clear indications of effects due to vibrational quanta.  We
believe that the observed values of the current are too small to
be attributed to shuttling (recall that the shuttling current has
the universal value of $1/2\pi$), and that the system is in the
tunneling limit.  In the experiment of Erbe {\it et
al.}\cite{erb-prl-01} the system was driven: indeed electrons were
shuttled, but the experiment was not designed to observe the
shuttling transition as a function of a control parameter.  Our
optimism is based on a number of new structures that are currently
being explored in the literature, and in particular we find the
structures of Scheible {\it et al.}\cite{sch-apl-04}, where a soft
silicon pillar forms the movable part, very promising.  We also
believe that the theoretical methods outlined above are suitable
for modeling quantitatively the forthcoming experiments.

\section{Conclusion}

We have discussed at length some properties of a nanomechanical
device, the quantum shuttle, which we believe can be an important
component in future applications, for example measurements of very
small displacements.  While the quantitative results of the
present work apply to very specific, and strongly idealized
models, we believe that many of the phenomena we address are
generic, and will be observed in near future.  One of the central
messages we want to pass is that the fluctuation properties of
these devices contain a wealth of information, and that this
information may be essential in identifying the key charge
transfer processes in these devices.


\begin{thebibliography}{41}
\expandafter\ifx\csname
natexlab\endcsname\relax\def\natexlab#1{#1}\fi
\expandafter\ifx\csname bibnamefont\endcsname\relax
  \def\bibnamefont#1{#1}\fi
\expandafter\ifx\csname bibfnamefont\endcsname\relax
  \def\bibfnamefont#1{#1}\fi
\expandafter\ifx\csname citenamefont\endcsname\relax
  \def\citenamefont#1{#1}\fi
\expandafter\ifx\csname url\endcsname\relax
  \def\url#1{\texttt{#1}}\fi
\expandafter\ifx\csname
urlprefix\endcsname\relax\def\urlprefix{URL }\fi
\providecommand{\bibinfo}[2]{#2}
\providecommand{\eprint}[2][]{\url{#2}}

\bibitem[{\citenamefont{Craighead}(2000)}]{cra-sci-00}
\bibinfo{author}{\bibfnamefont{H.~G.} \bibnamefont{Craighead}},
  \bibinfo{journal}{Science} \textbf{\bibinfo{volume}{290}},
  \bibinfo{pages}{1532} (\bibinfo{year}{2000}).

\bibitem[{\citenamefont{Cleland}(2003)}]{cleland}
\bibinfo{author}{\bibfnamefont{A.~N.} \bibnamefont{Cleland}},
  \emph{\bibinfo{title}{Foundations of Nanomechanics}}, Advanced Texts in
  Physics (\bibinfo{publisher}{Springer}, \bibinfo{address}{Berlin},
  \bibinfo{year}{2003}).

\bibitem[{\citenamefont{Blencowe}(2004)}]{ble-phr-04}
\bibinfo{author}{\bibfnamefont{M.}~\bibnamefont{Blencowe}},
  \bibinfo{journal}{Physics Reports} \textbf{\bibinfo{volume}{395}},
  \bibinfo{pages}{159} (\bibinfo{year}{2004}).

\bibitem[{\citenamefont{Gorelik et~al.}(1998)\citenamefont{Gorelik, Isacsson,
  Voinova, Kasemo, Shekhter, and Jonson}}]{gor-prl-98}
\bibinfo{author}{\bibfnamefont{L.~Y.} \bibnamefont{Gorelik}},
  \bibinfo{author}{\bibfnamefont{A.}~\bibnamefont{Isacsson}},
  \bibinfo{author}{\bibfnamefont{M.~V.} \bibnamefont{Voinova}},
  \bibinfo{author}{\bibfnamefont{B.}~\bibnamefont{Kasemo}},
  \bibinfo{author}{\bibfnamefont{R.~I.} \bibnamefont{Shekhter}},
  \bibnamefont{and} \bibinfo{author}{\bibfnamefont{M.}~\bibnamefont{Jonson}},
  \bibinfo{journal}{Phys. Rev. Lett.} \textbf{\bibinfo{volume}{80}},
  \bibinfo{pages}{4526} (\bibinfo{year}{1998}), \eprint{cond-mat/9711196}.

\bibitem[{\citenamefont{Park et~al.}(2000)\citenamefont{Park, Park, Lim,
  Anderson, Alivisatos, and McEuen}}]{par-nat-00}
\bibinfo{author}{\bibfnamefont{H.}~\bibnamefont{Park}},
  \bibinfo{author}{\bibfnamefont{J.}~\bibnamefont{Park}},
  \bibinfo{author}{\bibfnamefont{A.~K.~L.} \bibnamefont{Lim}},
  \bibinfo{author}{\bibfnamefont{E.~H.} \bibnamefont{Anderson}},
  \bibinfo{author}{\bibfnamefont{A.~P.} \bibnamefont{Alivisatos}},
  \bibnamefont{and} \bibinfo{author}{\bibfnamefont{P.~L.}
  \bibnamefont{McEuen}}, \bibinfo{journal}{Nature}
  \textbf{\bibinfo{volume}{407}}, \bibinfo{pages}{57} (\bibinfo{year}{2000}).

\bibitem[{\citenamefont{Boese and Schoeller}(2001)}]{boe-epl-01}
\bibinfo{author}{\bibfnamefont{D.}~\bibnamefont{Boese}} \bibnamefont{and}
  \bibinfo{author}{\bibfnamefont{H.}~\bibnamefont{Schoeller}},
  \bibinfo{journal}{Europhys. Lett.} \textbf{\bibinfo{volume}{54}},
  \bibinfo{pages}{668} (\bibinfo{year}{2001}), \eprint{cond-mat/0012140}.

\bibitem[{\citenamefont{McCarthy et~al.}(2003)\citenamefont{McCarthy,
  Prokof'ev, and Tuominen}}]{mcc-prb-03}
\bibinfo{author}{\bibfnamefont{K.~D.} \bibnamefont{McCarthy}},
  \bibinfo{author}{\bibfnamefont{N.}~\bibnamefont{Prokof'ev}},
  \bibnamefont{and} \bibinfo{author}{\bibfnamefont{M.~T.}
  \bibnamefont{Tuominen}}, \bibinfo{journal}{Phys. Rev. B}
  \textbf{\bibinfo{volume}{67}}, \bibinfo{pages}{245415}
  (\bibinfo{year}{2003}), \eprint{cond-mat/0205419}.

\bibitem[{\citenamefont{Flensberg}(2003)}]{fle-prb-03}
\bibinfo{author}{\bibfnamefont{K.}~\bibnamefont{Flensberg}},
  \bibinfo{journal}{Phys. Rev. B} \textbf{\bibinfo{volume}{68}},
  \bibinfo{pages}{205323} (\bibinfo{year}{2003}), \eprint{cond-mat/0302193}.

\bibitem[{\citenamefont{Braig and Flensberg}(2003)}]{braig-prb-03}
\bibinfo{author}{\bibfnamefont{S.}~\bibnamefont{Braig}} \bibnamefont{and}
  \bibinfo{author}{\bibfnamefont{K.}~\bibnamefont{Flensberg}},
  \bibinfo{journal}{Phys. Rev. B} \textbf{\bibinfo{volume}{68}},
  \bibinfo{pages}{205324} (\bibinfo{year}{2003}), \eprint{cond-mat/0303236}.

\bibitem[{\citenamefont{Fedorets et~al.}(2002)\citenamefont{Fedorets, Gorelik,
  Shekhter, and Jonson}}]{fed-epl-02}
\bibinfo{author}{\bibfnamefont{D.}~\bibnamefont{Fedorets}},
  \bibinfo{author}{\bibfnamefont{L.~Y.} \bibnamefont{Gorelik}},
  \bibinfo{author}{\bibfnamefont{R.~I.} \bibnamefont{Shekhter}},
  \bibnamefont{and} \bibinfo{author}{\bibfnamefont{M.}~\bibnamefont{Jonson}},
  \bibinfo{journal}{Europhys. Lett.} \textbf{\bibinfo{volume}{58}},
  \bibinfo{pages}{99} (\bibinfo{year}{2002}), \eprint{cond-mat/0104200}.

\bibitem[{\citenamefont{Kogan}(1996)}]{kogan}
\bibinfo{author}{\bibfnamefont{S.}~\bibnamefont{Kogan}},
  \emph{\bibinfo{title}{Electronic Noise and Fluctuations in Solids}}
  (\bibinfo{publisher}{Cambridge University Press},
  \bibinfo{address}{Cambridge}, \bibinfo{year}{1996}).

\bibitem[{\citenamefont{Blanter and B{\"{u}}ttiker}(2000)}]{bla-phr-00}
\bibinfo{author}{\bibfnamefont{Y.~M.} \bibnamefont{Blanter}} \bibnamefont{and}
  \bibinfo{author}{\bibfnamefont{M.}~\bibnamefont{B{\"{u}}ttiker}},
  \bibinfo{journal}{Physics Reports} \textbf{\bibinfo{volume}{336}},
  \bibinfo{pages}{1} (\bibinfo{year}{2000}).

\bibitem[{\citenamefont{Beenakker and Sch{\"{o}}nenberger}(2003)}]{bee-pht-03}
\bibinfo{author}{\bibfnamefont{C.}~\bibnamefont{Beenakker}} \bibnamefont{and}
  \bibinfo{author}{\bibfnamefont{C.}~\bibnamefont{Sch{\"{o}}nenberger}},
  \bibinfo{journal}{Physics Today} \textbf{\bibinfo{volume}{56}},
  \bibinfo{pages}{37} (\bibinfo{year}{2003}).

\bibitem[{\citenamefont{Mitra et~al.}(2003)\citenamefont{Mitra, Aleiner, and
  Millis}}]{mit-preprint-03}
\bibinfo{author}{\bibfnamefont{A.}~\bibnamefont{Mitra}},
  \bibinfo{author}{\bibfnamefont{I.}~\bibnamefont{Aleiner}}, \bibnamefont{and}
  \bibinfo{author}{\bibfnamefont{A.~J.} \bibnamefont{Millis}},
  \emph{\bibinfo{title}{Phonon effects in molecular transistors: Quantum and
  classical treatment}} (\bibinfo{year}{2003}), \eprint{cond-mat/0311503}.

\bibitem[{\citenamefont{Pistolesi}(2004)}]{pis-prb-04}
\bibinfo{author}{\bibfnamefont{F.}~\bibnamefont{Pistolesi}},
  \bibinfo{journal}{Phys. Rev. B} \textbf{\bibinfo{volume}{69}},
  \bibinfo{pages}{245409} (\bibinfo{year}{2004}), \eprint{cond-mat/0401361}.

\bibitem[{\citenamefont{Armour}(2004)}]{arm-preprint-04}
\bibinfo{author}{\bibfnamefont{A.~D.} \bibnamefont{Armour}},
  \emph{\bibinfo{title}{Current noise of a single-electron transistor coupled
  to a nano-mechanical resonator}} (\bibinfo{year}{2004}),
  \eprint{cond-mat/0401387}.

\bibitem[{\citenamefont{Chtchelkatchev
  et~al.}(2004)\citenamefont{Chtchelkatchev, Belzig, and
  Bruder}}]{cht-preprint-04}
\bibinfo{author}{\bibfnamefont{N.~M.} \bibnamefont{Chtchelkatchev}},
  \bibinfo{author}{\bibfnamefont{W.}~\bibnamefont{Belzig}}, \bibnamefont{and}
  \bibinfo{author}{\bibfnamefont{C.}~\bibnamefont{Bruder}},
  \emph{\bibinfo{title}{Charge transport through a set with a mechanically
  oscillating island}} (\bibinfo{year}{2004}), \eprint{cond-mat/0401486}.

\bibitem[{\citenamefont{Isacsson and Nord}(2004)}]{isa-epl-04}
\bibinfo{author}{\bibfnamefont{A.}~\bibnamefont{Isacsson}} \bibnamefont{and}
  \bibinfo{author}{\bibfnamefont{T.}~\bibnamefont{Nord}},
  \bibinfo{journal}{Europhys. Lett.} \textbf{\bibinfo{volume}{66}},
  \bibinfo{pages}{708} (\bibinfo{year}{2004}), \eprint{cond-mat/0402228}.

\bibitem[{\citenamefont{Novotn\'{y} et~al.}(2004)\citenamefont{Novotn\'{y},
  Donarini, Flindt, and Jauho}}]{nov-prl-04}
\bibinfo{author}{\bibfnamefont{T.}~\bibnamefont{Novotn\'{y}}},
  \bibinfo{author}{\bibfnamefont{A.}~\bibnamefont{Donarini}},
  \bibinfo{author}{\bibfnamefont{C.}~\bibnamefont{Flindt}}, \bibnamefont{and}
  \bibinfo{author}{\bibfnamefont{A.-P.} \bibnamefont{Jauho}},
  \bibinfo{journal}{Phys. Rev. Lett.} \textbf{\bibinfo{volume}{92}},
  \bibinfo{pages}{248302} (\bibinfo{year}{2004}), \eprint{cond-mat/0402597}.

\bibitem[{\citenamefont{Blanter et~al.}(2004)\citenamefont{Blanter, Usmani, and
  Nazarov}}]{bla-prl-04}
\bibinfo{author}{\bibfnamefont{Y.~M.} \bibnamefont{Blanter}},
  \bibinfo{author}{\bibfnamefont{O.}~\bibnamefont{Usmani}}, \bibnamefont{and}
  \bibinfo{author}{\bibfnamefont{Y.~V.} \bibnamefont{Nazarov}},
  \bibinfo{journal}{Phys. Rev. Lett.} \textbf{\bibinfo{volume}{93}},
  \bibinfo{pages}{136802} (\bibinfo{year}{2004}), \eprint{cond-mat/0404615}.

\bibitem[{\citenamefont{Reulet et~al.}(2004)\citenamefont{Reulet, Senzier, and
  Prober}}]{reu-prl-03}
\bibinfo{author}{\bibfnamefont{B.}~\bibnamefont{Reulet}},
  \bibinfo{author}{\bibfnamefont{J.}~\bibnamefont{Senzier}}, \bibnamefont{and}
  \bibinfo{author}{\bibfnamefont{D.~E.} \bibnamefont{Prober}},
  \bibinfo{journal}{Phys. Rev. Lett.} \textbf{\bibinfo{volume}{91}},
  \bibinfo{pages}{196601} (\bibinfo{year}{2004}).

\bibitem[{\citenamefont{Flindt et~al.}(2004{\natexlab{a}})\citenamefont{Flindt,
  Novotn\'{y}, and Jauho}}]{fli-epl-04}
\bibinfo{author}{\bibfnamefont{C.}~\bibnamefont{Flindt}},
  \bibinfo{author}{\bibfnamefont{T.}~\bibnamefont{Novotn\'{y}}},
  \bibnamefont{and} \bibinfo{author}{\bibfnamefont{A.-P.} \bibnamefont{Jauho}},
  \emph{\bibinfo{title}{Full counting statistics of nano-electromechanical
  systems}} (\bibinfo{year}{2004}{\natexlab{a}}), \eprint{cond-mat/0410322}.

\bibitem[{\citenamefont{Novotn\'{y} et~al.}(2003)\citenamefont{Novotn\'{y},
  Donarini, and Jauho}}]{nov-prl-03}
\bibinfo{author}{\bibfnamefont{T.}~\bibnamefont{Novotn\'{y}}},
  \bibinfo{author}{\bibfnamefont{A.}~\bibnamefont{Donarini}}, \bibnamefont{and}
  \bibinfo{author}{\bibfnamefont{A.-P.} \bibnamefont{Jauho}},
  \bibinfo{journal}{Phys. Rev. Lett.} \textbf{\bibinfo{volume}{90}},
  \bibinfo{pages}{256801} (\bibinfo{year}{2003}), \eprint{cond-mat/0301441}.

\bibitem[{\citenamefont{Donarini et~al.}(2004)\citenamefont{Donarini,
  Novotn\'{y}, and Jauho}}]{modena}
\bibinfo{author}{\bibfnamefont{A.}~\bibnamefont{Donarini}},
  \bibinfo{author}{\bibfnamefont{T.}~\bibnamefont{Novotn\'{y}}},
  \bibnamefont{and} \bibinfo{author}{\bibfnamefont{A.-P.} \bibnamefont{Jauho}},
  \bibinfo{journal}{Semicond. Sci. Technol.} \textbf{\bibinfo{volume}{19}},
  \bibinfo{pages}{S430} (\bibinfo{year}{2004}), \eprint{cond-mat/0401357}.

\bibitem[{\citenamefont{Flindt et~al.}(2004{\natexlab{b}})\citenamefont{Flindt,
  Novotn\'{y}, and Jauho}}]{fli-prb-04}
\bibinfo{author}{\bibfnamefont{C.}~\bibnamefont{Flindt}},
  \bibinfo{author}{\bibfnamefont{T.}~\bibnamefont{Novotn\'{y}}},
  \bibnamefont{and} \bibinfo{author}{\bibfnamefont{A.-P.} \bibnamefont{Jauho}},
  \bibinfo{journal}{Phys. Rev. B}  (\bibinfo{year}{2004}{\natexlab{b}}),
  \eprint{cond-mat/0405512}.

\bibitem[{\citenamefont{Donarini}(2004)}]{andrea}
\bibinfo{author}{\bibfnamefont{A.}~\bibnamefont{Donarini}}, Ph.D. thesis,
  \bibinfo{school}{MIC, Technical University of Denmark}
  (\bibinfo{year}{2004}),
  \urlprefix\url{http://www.mic.dtu.dk/research/TheoreticalNano/publications/T%
heses.htm}.

\bibitem[{\citenamefont{Shekhter et~al.}(2003)\citenamefont{Shekhter, Galperin,
  Gorelik, Isacsson, and Jonson}}]{she-jpc-03}
\bibinfo{author}{\bibfnamefont{R.~I.} \bibnamefont{Shekhter}},
  \bibinfo{author}{\bibfnamefont{Y.}~\bibnamefont{Galperin}},
  \bibinfo{author}{\bibfnamefont{L.~Y.} \bibnamefont{Gorelik}},
  \bibinfo{author}{\bibfnamefont{A.}~\bibnamefont{Isacsson}}, \bibnamefont{and}
  \bibinfo{author}{\bibfnamefont{M.}~\bibnamefont{Jonson}},
  \bibinfo{journal}{J. Phys. Condens. Matter} \textbf{\bibinfo{volume}{15}},
  \bibinfo{pages}{R441} (\bibinfo{year}{2003}).

\bibitem[{\citenamefont{Fedorets}(2003)}]{fed-prb-03}
\bibinfo{author}{\bibfnamefont{D.}~\bibnamefont{Fedorets}},
  \bibinfo{journal}{Phys. Rev. B} \textbf{\bibinfo{volume}{68}},
  \bibinfo{pages}{033106} (\bibinfo{year}{2003}).

\bibitem[{\citenamefont{Fedorets
  et~al.}(2004{\natexlab{a}})\citenamefont{Fedorets, Gorelik, Shekhter, and
  Jonson}}]{fed-prl-04}
\bibinfo{author}{\bibfnamefont{D.}~\bibnamefont{Fedorets}},
  \bibinfo{author}{\bibfnamefont{L.~Y.} \bibnamefont{Gorelik}},
  \bibinfo{author}{\bibfnamefont{R.~I.} \bibnamefont{Shekhter}},
  \bibnamefont{and} \bibinfo{author}{\bibfnamefont{M.}~\bibnamefont{Jonson}},
  \bibinfo{journal}{Phys. Rev. Lett.} \textbf{\bibinfo{volume}{92}},
  \bibinfo{pages}{166801} (\bibinfo{year}{2004}{\natexlab{a}}),
  \eprint{cond-mat/0311105}.

\bibitem[{\citenamefont{Smirnov et~al.}(2004)\citenamefont{Smirnov, Mourokh,
  and Horing}}]{smi-prb-04}
\bibinfo{author}{\bibfnamefont{A.~Y.} \bibnamefont{Smirnov}},
  \bibinfo{author}{\bibfnamefont{L.~G.} \bibnamefont{Mourokh}},
  \bibnamefont{and} \bibinfo{author}{\bibfnamefont{N.~J.~M.}
  \bibnamefont{Horing}}, \bibinfo{journal}{Phys. Rev. B}
  \textbf{\bibinfo{volume}{69}}, \bibinfo{pages}{155310}
  (\bibinfo{year}{2004}), \eprint{cond-mat/0311052}.

\bibitem[{\citenamefont{Armour and MacKinnon}(2002)}]{arm-prb-02}
\bibinfo{author}{\bibfnamefont{A.~D.} \bibnamefont{Armour}} \bibnamefont{and}
  \bibinfo{author}{\bibfnamefont{A.}~\bibnamefont{MacKinnon}},
  \bibinfo{journal}{Phys. Rev. B} \textbf{\bibinfo{volume}{66}},
  \bibinfo{pages}{035333} (\bibinfo{year}{2002}), \eprint{cond-mat/0204521}.

\bibitem[{\citenamefont{Gardiner and Zoller}(2000)}]{gardiner}
\bibinfo{author}{\bibfnamefont{C.~W.} \bibnamefont{Gardiner}} \bibnamefont{and}
  \bibinfo{author}{\bibfnamefont{P.}~\bibnamefont{Zoller}},
  \emph{\bibinfo{title}{Quantum Noise}} (\bibinfo{publisher}{Springer},
  \bibinfo{year}{2000}), \bibinfo{edition}{2nd} ed.

\bibitem[{\citenamefont{Weiss}(1999)}]{weiss}
\bibinfo{author}{\bibfnamefont{U.}~\bibnamefont{Weiss}},
  \emph{\bibinfo{title}{Quantum Dissipative Systems}},
  vol.~\bibinfo{volume}{10} of \emph{\bibinfo{series}{Series in Modern
  Condensed Matter Physics}} (\bibinfo{publisher}{World Scientific},
  \bibinfo{year}{1999}), \bibinfo{edition}{2nd} ed.

\bibitem[{\citenamefont{Gurvitz and Prager}(1996)}]{gur-prb-96}
\bibinfo{author}{\bibfnamefont{S.~A.} \bibnamefont{Gurvitz}} \bibnamefont{and}
  \bibinfo{author}{\bibfnamefont{Y.~S.} \bibnamefont{Prager}},
  \bibinfo{journal}{Phys. Rev. B} \textbf{\bibinfo{volume}{53}},
  \bibinfo{pages}{15932} (\bibinfo{year}{1996}).

\bibitem[{\citenamefont{Elattari and Gurvitz}(2002)}]{ela-pla-02}
\bibinfo{author}{\bibfnamefont{B.}~\bibnamefont{Elattari}} \bibnamefont{and}
  \bibinfo{author}{\bibfnamefont{S.~A.} \bibnamefont{Gurvitz}},
  \bibinfo{journal}{Physics Letters A} \textbf{\bibinfo{volume}{292}},
  \bibinfo{pages}{289} (\bibinfo{year}{2002}).

\bibitem[{\citenamefont{Flindt}(2004)}]{christian}
\bibinfo{author}{\bibfnamefont{C.}~\bibnamefont{Flindt}}, Master's thesis,
  \bibinfo{school}{MIC, Technical University of Denmark}
  (\bibinfo{year}{2004}),
  \urlprefix\url{http://www.mic.dtu.dk/research/TheoreticalNano/publications/T%
heses.htm}.

\bibitem[{\citenamefont{Fedorets
  et~al.}(2004{\natexlab{b}})\citenamefont{Fedorets, Gorelik, Shekhter, and
  Jonson}}]{fed-preprint-04}
\bibinfo{author}{\bibfnamefont{D.}~\bibnamefont{Fedorets}},
  \bibinfo{author}{\bibfnamefont{L.~Y.} \bibnamefont{Gorelik}},
  \bibinfo{author}{\bibfnamefont{R.~I.} \bibnamefont{Shekhter}},
  \bibnamefont{and} \bibinfo{author}{\bibfnamefont{M.}~\bibnamefont{Jonson}},
  \emph{\bibinfo{title}{Spintronics of a nanoelectromechanical shuttle}}
  (\bibinfo{year}{2004}{\natexlab{b}}), \eprint{cond-mat/0408591}.

\bibitem[{\citenamefont{Jordan and Sukhorukov}(2004)}]{jor-preprint-04}
\bibinfo{author}{\bibfnamefont{A.~N.} \bibnamefont{Jordan}} \bibnamefont{and}
  \bibinfo{author}{\bibfnamefont{E.~V.} \bibnamefont{Sukhorukov}},
  \emph{\bibinfo{title}{Transport statistics of bistable systems}}
  (\bibinfo{year}{2004}), \eprint{cond-mat/0406261}.

\bibitem[{\citenamefont{Kie{\ss}lich et~al.}(2003)\citenamefont{Kie{\ss}lich,
  Wacker, and Sch{\"{o}}ll}}]{Kie-prb-03}
\bibinfo{author}{\bibfnamefont{G.}~\bibnamefont{Kie{\ss}lich}},
  \bibinfo{author}{\bibfnamefont{A.}~\bibnamefont{Wacker}}, \bibnamefont{and}
  \bibinfo{author}{\bibfnamefont{E.}~\bibnamefont{Sch{\"{o}}ll}},
  \bibinfo{journal}{Phys. Rev. B} \textbf{\bibinfo{volume}{68}},
  \bibinfo{pages}{125320} (\bibinfo{year}{2003}), \eprint{cond-mat/0303025}.

\bibitem[{\citenamefont{Erbe et~al.}(2001)\citenamefont{Erbe, Weiss, Zwerger,
  and Blick}}]{erb-prl-01}
\bibinfo{author}{\bibfnamefont{A.}~\bibnamefont{Erbe}},
  \bibinfo{author}{\bibfnamefont{C.}~\bibnamefont{Weiss}},
  \bibinfo{author}{\bibfnamefont{W.}~\bibnamefont{Zwerger}}, \bibnamefont{and}
  \bibinfo{author}{\bibfnamefont{R.~H.} \bibnamefont{Blick}},
  \bibinfo{journal}{Phys. Rev. Lett.} \textbf{\bibinfo{volume}{87}},
  \bibinfo{pages}{096106} (\bibinfo{year}{2001}), \eprint{cond-mat/0011429}.

\bibitem[{\citenamefont{Scheible and Blick}(2004)}]{sch-apl-04}
\bibinfo{author}{\bibfnamefont{D.~C.} \bibnamefont{Scheible}} \bibnamefont{and}
  \bibinfo{author}{\bibfnamefont{R.~H.} \bibnamefont{Blick}},
  \bibinfo{journal}{Appl. Phys. Lett.} \textbf{\bibinfo{volume}{84}},
  \bibinfo{pages}{4632} (\bibinfo{year}{2004}).

\end{thebibliography}

\end{document}